# A Simple Test on 2-Vertex- and 2-Edge-Connectivity

Jens M. Schmidt
MPI für Informatik, Saarbrücken
(jens.schmidt@mpi-inf.mpg.de)

**Abstract**

Testing a graph on 2-vertex- and 2-edge-connectivity are two fundamental algorithmic graph problems. For both problems, different linear-time algorithms with simple implementations are known. Here, an even simpler linear-time algorithm is presented that computes a structure from which both the 2-vertex- and 2-edge-connectivity of a graph can be easily "read off". The algorithm computes all bridges and cut vertices of the input graph in the same time.

## 1 Introduction

Testing a graph on 2-connectivity (i. e., 2-vertex-connectivity) and on 2-edge-connectivity are fundamental algorithmic graph problems. Tarjan presented the first linear-time algorithms for these problems, respectively [11, 12]. Since then, many linear-time algorithms have been given (e. g., [2, 3, 4, 5, 6, 13, 14, 15]) that compute structures which inherently characterize either the 2- or 2-edge-connectivity of a graph. Examples include *open ear decompositions* [8, 16], *block-cut trees* [7], *bipolar orientations* [2] and *s-t-numberings* [2] (all of which can be used to determine 2-connectivity) and *ear decompositions* [8] (the existence of which determines 2-edge-connectivity).

Most of the mentioned algorithms use a depth-first search-tree (DFS-tree) and compute so-called *low-point* values, which are defined in terms of a DFS-tree (see [11] for a definition of low-points). This is a concept Tarjan introduced in his first algorithms and that has been applied successfully to many graph problems later on. However, low-points do not always provide the most natural solution: Brandes [2] and Gabow [6] gave considerably simpler algorithms for computing most of the above-mentioned structures (and testing 2-connectivity) by using simple path-generating rules instead of low-points; they call these algorithms *path-based*.

The aim of this paper is a self-contained exposition of an even simpler linear-time algorithm that tests both the 2- and 2-edge-connectivity of a graph. It is suitable for teaching in introductory courses on algorithms. While Tarjan's two algorithms are currently the most popular ones used for teaching (see [6] for a list of 11 text books in which they appear), in my teaching experience, undergraduate students have difficulties with the details of using low-points.

The algorithm presented here uses a very natural path-based approach instead of low-points; similar approaches have been presented by Ramachandran [10] and Tsin [14] in the context of parallel and distributed algorithms, respectively. The approach is related to ear decompositions; in fact, it computes an (open) ear decomposition if the input graph has appropriate connectivity.

**Notation.** We use standard graph-theoretic terminology from [1]. Let $\delta(G)$ be the minimum degree of a graph $G$. A *cut vertex* is a vertex in a connected graph that disconnects the graph upon deletion. Similarly, a *bridge* is an edge in a connected graph that disconnects the graph upon deletion. A graph is 2-connected if it is connected and contains at least 3 vertices, but no cut vertex. A graph is 2-edge-connected if it is connected and contains at least 2 vertices, but no bridge. Note that for very small graphs, different definitions of (edge)connectivity are used in literature; here, we chose the common definition that ensures consistency with Menger's Theorem [9]. It is easy to see that every 2-connected graph is 2-edge-connected, as otherwise any bridge in this graph on at least 3 vertices would have an end point that is a cut vertex.

## 2 Decomposition into Chains

We will decompose the input graph into a set of paths and cycles, each of which will be called a *chain*. Some easy-to-check properties on these chains will then characterize both the 2- and 2-edge-connectivity of the graph. Let $G = (V, E)$ be the input graph and assume for convenience that $G$ is simple and that $|V| \geq 3$.

We first perform a depth-first search on $G$. This implicitly checks $G$ on being connected. If $G$ is connected, we get a DFS-tree $T$ that is rooted on a vertex $r$; otherwise, we stop, as $G$ is neither 2- nor 2-edge-connected. The DFS assigns a *depth-first index* (DFI) to every vertex. We assume that all *tree edges* (i.e., edges in $T$) are oriented towards $r$ and all *backedges* (i.e., edges that are in $G$ but not in $T$) are oriented away from $r$. Thus, every backedge lies in exactly one *directed cycle* $C(e)$. Let every vertex be marked as *unvisited*.

We now decompose $G$ into *chains* by applying the following procedure for each vertex $v$ in ascending DFI-order: For every backedge $e$ that starts at $v$, we traverse $C(e)$, beginning with $v$, and stop at the first vertex that is marked as visited. During such a traversal, every traversed vertex is marked as *visited*. Thus, a traversal stops at the latest at $v$ and forms either a directed path or cycle, beginning with $v$; we call this path or cycle a *chain* and identify it with the list of vertices and edges in the order in which they were visited. The $i$th chain found by this procedure is referred to as $C_i$.

The chain $C_1$, if exists, is a cycle, as every vertex is unvisited at the beginning (note $C_1$ does not have to contain $r$). There are $|E| - |V| + 1$ chains, as every of the $|E| - |V| + 1$ backedges creates exactly one chain. We call the set $C = \{C_1, \ldots, C_{|E|-|V|+1}\}$ a *chain decomposition*; see Figure 1 for an example.

Clearly, a chain decomposition can be computed in linear time. This almost concludes the algorithmic part; we now state easy-to-check conditions on $C$ that characterize 2- and 2-edge-connectivity. All proofs will be given in the next section.

**Theorem 1.** *Let $C$ be a chain decomposition of a simple connected graph $G$. Then $G$ is 2-edge-connected if and only if the chains in $C$ partition $E$.*

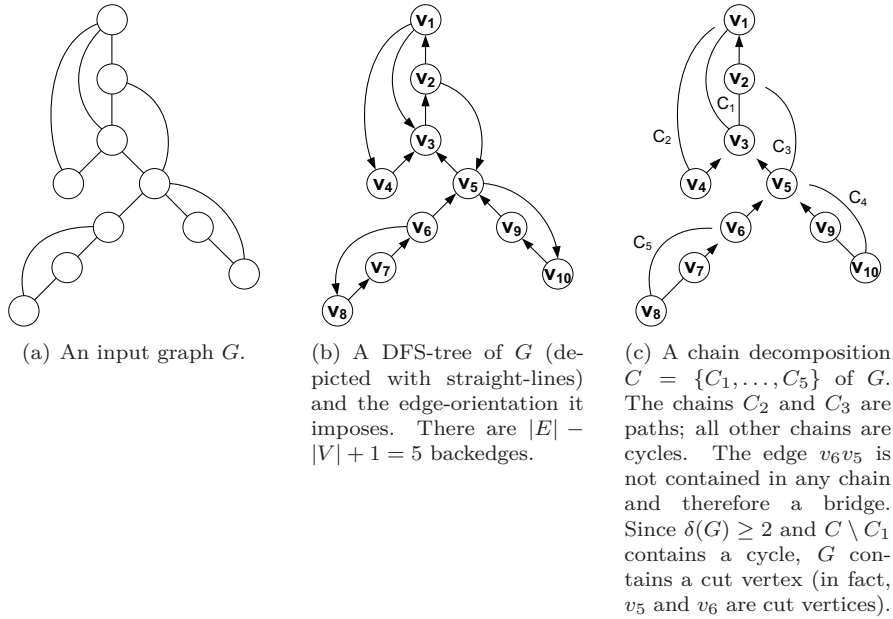

(a) An input graph $G$.

(b) A DFS-tree of $G$ (depicted with straight-lines) and the edge-orientation it imposes. There are $|E| - |V| + 1 = 5$ backedges.

(c) A chain decomposition $C = \{C_1, \ldots, C_5\}$ of $G$. The chains $C_2$ and $C_3$ are paths; all other chains are cycles. The edge $v_6 v_5$ is not contained in any chain and therefore a bridge. Since $\delta(G) \geq 2$ and $C \setminus C_1$ contains a cycle, $G$ contains a cut vertex (in fact, $v_5$ and $v_6$ are cut vertices).

Figure 1: A graph $G$, its DFS-tree and a chain decomposition of $G$.

**Theorem 2.** *Let $C$ be a chain decomposition of a simple 2-edge-connected graph $G$. Then $G$ is 2-connected if and only if $C_1$ is the only cycle in $C$.*

The properties in Theorems 1 and 2 can be efficiently tested: In order to check whether $C$ partitions $E$, we mark every edge that is traversed by the chain decomposition. If $G$ is 2-edge-connected, every $C_i$ can be checked on forming a cycle by comparing its first and last vertex on identity. For pseudo-code, see Algorithm 1.

---
**Algorithm 1** Check(graph $G$)  ▷ $G$ is simple and connected with $|V| \geq 3$
---
1: Compute a DFS-tree $T$ of $G$
2: Compute a chain decomposition $C$; mark every visited edge
3: **if** $G$ contains an unvisited edge **then**
4:     output "NOT 2-EDGE-CONNECTED"
5: **else if** there is a cycle in $C$ different from $C_1$ **then**
6:     output "2-EDGE-CONNECTED BUT NOT 2-CONNECTED"
7: **else**
8:     output "2-CONNECTED"
---

We state a variant of Theorem 2, which does not rely on edge-connectivity. Its proof is very similar to the one of Theorem 2.

**Theorem 3.** *Let $C$ be a chain decomposition of a simple connected graph $G$. Then $G$ is 2-connected if and only if $\delta(G) \geq 2$ and $C_1$ is the only cycle in $C$.*

## 3 Proofs

It remains to give the proofs of Theorems 1 and 2. For a tree $T$ rooted at $r$ and a vertex $x$ in $T$, let $T(x)$ be the subtree of $T$ that consists of $x$ and all descendants of $x$ (independent of the edge orientations of $T$). Theorem 1 is immediately implied by the following lemma.

**Lemma 4.** *Let $C$ be a chain decomposition of a simple connected graph $G$. An edge $e$ in $G$ is a bridge if and only if $e$ is not contained in any chain in $C$.*

*Proof.* Let $e$ be a bridge and assume to the contrary that $e$ is contained in a chain whose first edge (i.e., whose backedge) is $b$. The bridge $e$ is not contained in any cycle of $G$, as otherwise the end points of $e$ would still be connected when deleting $e$, contradicting that $e$ is a bridge. This contradicts the fact that $e$ is contained in the cycle $C(b)$.

Let $e$ be an edge that is not contained in any chain in $C$. Let $T$ be the DFS-tree that was used for computing $C$ and let $x$ be the end point of $e$ that is farthest away from the root $r$ of $T$, in particular $x \neq r$. Then $e$ is a tree-edge, as otherwise $e$ would be contained in a chain. For the same reason, there is no backedge with exactly one end point in $T(x)$. Deleting $e$ therefore disconnects all vertices in $T(x)$ from $r$. Hence, $e$ is a bridge. □

The following lemma implies Theorem 2, as every 2-edge-connected graph has minimum degree 2.

**Lemma 5.** *Let $C$ be a chain decomposition of a simple connected graph $G$ with $\delta(G) \geq 2$. A vertex $v$ in $G$ is a cut vertex if and only if $v$ is incident to a bridge or $v$ is the first vertex of a cycle in $C \setminus C_1$.*

*Proof.* Let $v$ be a cut vertex in $G$; we may assume that $v$ is not incident to a bridge. Let $X$ and $Y$ be connected components of $G \setminus v$. Then $X$ and $Y$ have to contain at least two neighbors of $v$ in $G$, respectively. Let $X^{+v}$ and $Y^{+v}$ denote the subgraphs of $G$ that are induced by $X \cup v$ and $Y \cup v$, respectively. Both $X^{+v}$ and $Y^{+v}$ contain a cycle through $v$, as both $X$ and $Y$ are connected. It follows that $C_1$ exists; assume w.l.o.g. that $C_1 \notin X^{+v}$. Then there is at least one backedge in $X^{+v}$ that starts at $v$. When the first such backedge is traversed in the chain decomposition, every vertex in $X$ is still unvisited. The traversal therefore closes a cycle that starts at $v$ and is different from $C_1$, as $C_1 \notin X^{+v}$.

If $v$ is incident to a bridge, $\delta(G) \geq 2$ implies that $v$ is a cut vertex. Let $v$ be the first vertex of a cycle $C_i \neq C_1$ in $C$. If $v$ is the root $r$ of the DFS-tree $T$ that was used for computing $C$, both cycles $C_1$ and $C_i$ end at $v$. Thus, $v$ has at least two children in $T$ and $v$ must be a cut vertex. Otherwise $v \neq r$; let $wv$ be the last edge in $C_i$. Then no backedge starts at a vertex with smaller DFI than $v$ and ends at a vertex in $T(w)$, as otherwise $vw$ would not be contained in $C_i$. Thus, deleting $v$ separates $r$ from all vertices in $T(w)$ and $v$ is a cut vertex. □

## 4 Extensions

We state how some additional structures can be computed from a chain decomposition. Note that Lemmas 4 and 5 can be used to compute all bridges and cut vertices of $G$ in linear time. Having these, the 2-*connected components* (i.e., the maximal 2-connected subgraphs) of $G$ and the 2-edge-connected components (i. e., the maximal 2-edge-connected subgraphs) of $G$ can be easily obtained. This gives the so-called *block-cut tree* [7] of $G$, which represents the dependence of the 2-connected components and cut vertices in $G$ in a tree (it gives also the corresponding tree representing the 2-edge-connected components and bridges of $G$).

Additionally, the set of chains $C$ computed by our algorithm is an ear decomposition if $G$ is 2-edge-connected and an open ear decomposition if $G$ is 2-connected. Note that $C$ is not an arbitrary (open) ear decomposition, as it depends on the DFS-tree. The existence of these ear decompositions characterize the 2-(edge-)connectivity of a graph [8, 16]; Brandes [2] gives a simple linear-time transformation that computes a bipolar orientation and a $s$-$t$-numbering from such an open ear decomposition.

# A  Appendix

We omitted the proof of Theorem 3, as it is very similar to the one of Theorem 2. For completeness, we give the proof here.

*Proof of Theorem 3:* Let $T$ be the DFS-tree that was used for computing $C$ and let $r$ be its root. First, let $G$ be 2-connected. Clearly, this implies $\delta(G) \geq 2$. Moreover, $r$ has exactly one child, as otherwise $r$ would be a cut vertex. Thus, $r$ is incident to a backedge, which implies that $C_1$ exists and is a cycle that starts at $r$. Assume to the contrary that $v$ is the first vertex of a cycle $C_i \neq C_1$. If $v = r$, both cycles $C_1$ and $C_i$ end at $v$. Thus, $v$ has at least two children in $T$. This implies that $v$ is a cut vertex, which contradicts the 2-connectivity of $G$. If $v \neq r$, let $wv$ be the last edge in $C_i$. There is no backedge that starts at a vertex with smaller DFI than $v$ and ends at a vertex in $T(w)$, as otherwise $wv$ would be contained in a chain $C_j$ with $j < i$. Thus, deleting $v$ disconnects $r$ from all vertices in $T(w)$, which contradicts the 2-connectivity of $G$.

Let $\delta(G) \geq 2$ and $C_1$ be the only cycle in $C$ and assume to the contrary that $G$ is not 2-connected. Then $G$ contains a cut vertex $v$, as $\delta(G) \geq 2$ implies $|V| \geq 3$. Clearly, $C_1$ can intersect with at most one connected component of $G \setminus v$. Let $X$ be a connected component of $G \setminus v$ that does not contain any vertex of $C_1$. Let $X^{+v}$ be the subgraph of $G$ that is induced by $X \cup v$. There must be a cycle in $X^{+v}$, as otherwise $X^{+v}$ would be a tree, whose leafs would contradict $\delta(G) \geq 2$. Hence, $X^{+v}$ contains at least one backedge; let $b$ be the first backedge in $X^{+v}$ that is traversed by the chain decomposition. As $r \notin X$, all vertices in $D(b)$ except the start point $w$ of $b$ have greater DFIs than $w$. Thus, the traversal on $b$ computes a chain $C_i \subset X^{+v}$ that is a cycle and that is distinct from $C_1$, as $X$ does not contain any vertex of $C_1$. □